\documentclass[12pt]{article}

\usepackage{amsmath}
\usepackage{graphicx}
\usepackage{hyperref}
\usepackage[FIGTOPCAP]{subfigure}

\begin{document}

\title{Small oscillations of a heavy symmetric top when magnitudes of conserved angular momenta are equal}

\author{        V. Tanr{\i}verdi \\
tanriverdivedat@googlemail.com \\
Address: Bahad{\i}n Kasabas{\i} 66710 Sorgun-Yozgat TURKEY
}

\date{}

\maketitle

\begin{abstract}
Small oscillations of a heavy symmetric top are studied when magnitudes of conserved angular momenta are equal to each other.
Results show that the small oscillation approximation can be used in these cases.
\end{abstract}

\section{Introduction}

Small oscillations of a heavy symmetric top are previously studied by different authors \cite{Crabtree, Lamb, Goldstein, Arnold, Greiner, Tanriverdi_so, Routh}.
One can use the results of the small oscillation approximation to study the motion of a heavy symmetric top.
This approximation can be very helpful in teaching.

It is known that there are some changes in motion when conserved angular momenta are equal to each other \cite{Routh, KleinSommerfeld, Symon, Gray, Tanriverdi_abeql}.
One can see from these works that the motion including upright position is only possible when the two conserved angular momenta are equal.
If one of the conserved angular momenta is equal to the negative of the other one, then the motion at the bottom position is possible.
These are necessary to explain some of the observed motions of a heavy symmetric top and can be helpful in teaching.

In this work, we will study the motion of a heavy symmetric top when conserved angular momenta are equal by using the small oscillation approximation.
In section \ref{one}, we will consider the motion when conserved angular momenta are equal with some small advancements.
In section \ref{two}, we will study small oscillations when conserved angular momenta are equal.
Then, we will conclude.
In the appendix, we will consider the case with a heavier counterweight.

\section{Heavy symmetric top when $|b|=|a|$}
\label{one}

In most of the works on heavy symmetric top, the situation when conserved angular momenta are different is studied.
In the previous work, we have studied small oscillations of a heavy symmetric top when $|b|\ne|a|$ by using \cite{Tanriverdi_so} 
\begin{equation}
	        \ddot \theta=\frac{\sin \theta}{I_x} \left[ Mgl+I_x \dot \phi^2 \cos \theta-I_z\dot \phi^2 \cos \theta-I_z \dot \phi \dot \psi \right]. \label{ddottheta}
\end{equation}
where $a=L_z/I_x$ and $b=L_{z'}/I_x$ are two constants defined from conserved angular momenta.
This equation is written in terms of Euler angles and their derivatives, and $I_x$ and $I_z$ are moments of inertia, $M$ is the mass, $l$ is the distance between the center of mass and $g$ is the gravitational acceleration.

When magnitudes of conserved angular momenta are equal to each other, which is equivalent to $|b|=|a|$, there are some changes, and at the following part, we will consider these cases.
We will consider $Mgl$ as positive unless stated otherwise.

\subsection{Heavy symmetric top when $b=a$}

When $b=a$, the effective potential can be written as \cite{Tanriverdi_abeql} 
\begin{equation}
        U_{eff}(\theta)= \frac{I_x}{2}\frac{a^2 (1-\cos \theta)}{1+\cos \theta}+Mgl \cos \theta.
        \label{ueff2}
\end{equation}
In this case, the infinity at $\theta=0$ disappears and the motion at the upright position becomes possible.
When the motion at $\theta=0$ is possible, one may need to extend the domain from $\theta \in [0,\pi]$ to $\theta \in [-\pi,\pi]$ to avoid the discontinuity in $\dot \theta$ \cite{Tanriverdi_abeql}.
We will use such extended domains for these types of motion.

\begin{figure}[h!]
\begin{center}
\subfigure[$b=a$ (fast top)]{
\includegraphics[width=4.0cm]{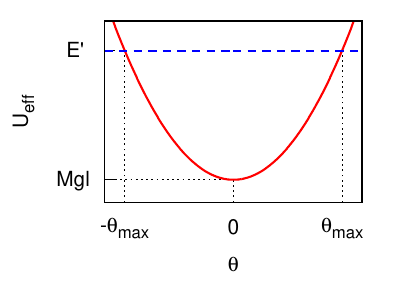}
}
\subfigure[$b=a$ (slow top)]{
\includegraphics[width=4.0cm]{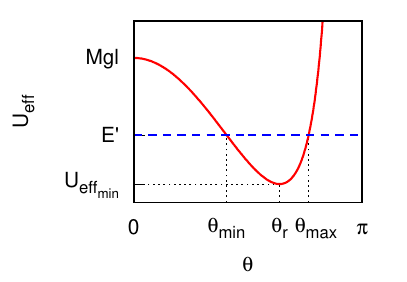}
}
\subfigure[$b=-a$]{
\includegraphics[width=4.0cm]{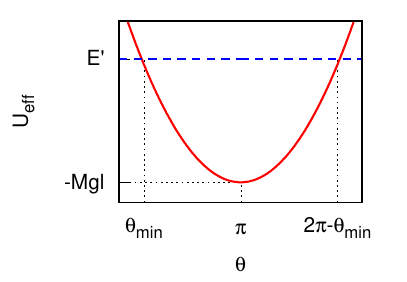}
}
\caption{
	Effective potentials.
        (a) For fast top when $b=a$ 
        (b) For slow top when $b=a$
        (c) When $b=-a$
        }
\label{fig:ueff}
\end{center}
\end{figure}

The derivative of the effective potential, equation \eqref{ueff2}, can be obtained as
\begin{equation}
\frac{d U_{eff}(\theta)}{d \theta}= \left[\frac{ I_x a^2 }{(1+\cos \theta)^2}-Mgl \right] \sin \theta.
\end{equation}
The terms in the parentheses can only be equal to zero when $|a|<\sqrt{4 Mgl/I_x}$.
When $|a|>\sqrt{4 Mgl/I_x}$, i.e. fast top, $\theta=0$ corresponds to the minimum which is equal to $Mgl$. 
When $|a|<\sqrt{4 Mgl/I_x}$, i.e. slow top, $\theta=0$ corresponds to the local maximum with value $Mgl$, 
and the minimum occurs at $\theta_r =\arccos(\sqrt{I_x a^2/Mgl}-1)$ with value $U_{eff_{min}}=2\sqrt{Mgl I_x a^2}-Mgl-I_x a^2/2$.
These two situations can be seen in figure \ref{fig:ueff}(a) and \ref{fig:ueff}(b). 

For slow top, turning angles can be found by considering $E'=U_{eff}$ with $\dot \theta=0$.
In this case, by changing varible $u=\cos \theta$, one gets a quadratic equation for $u$ as
\begin{equation}
	u^2+ \left(1-\frac{I_x a^2+2 E'}{2 Mgl}\right) u+\frac{I_x a^2-2 E'}{2 Mgl}=0, \label{quadratic_abeql}
\end{equation}
From the roots of this equation one can obtain turning angles by using $\theta_{min,max}=\arccos(u_{1,2})$, see figure \ref{fig:ueff}(b).

For fast tops, there exists only one root, i.e. $\theta_{max}$, between $0$ and $\pi$ which corresponds to turning angle,
and for the extended domain, the other turning angle can be taken as $-\theta_{max}$, see figure \ref{fig:ueff}(a).

When $b=a$, angular velocities $\dot \phi$ and $\dot \psi$ can be found as \cite{Tanriverdi_abeql}
\begin{eqnarray}
\dot \phi &=& \frac{a}{1+\cos \theta}, \\
\dot \psi &=& a\left( \frac{I_x}{I_z} -\frac{\cos \theta}{1+\cos \theta}\right).
\end{eqnarray}

\subsection{Heavy symmetric top when $b=-a$}

When $b=-a$, the effective potential can be written as \cite{Tanriverdi_abeql}
\begin{equation}
        U_{eff}(\theta)= \frac{I_x}{2}\frac{a^2 (1+\cos \theta)}{1-\cos \theta}+Mgl \cos \theta.
        \label{ueff3}
\end{equation}
In this case, the infinity at $\theta=\pi$ disappears, and the motion at there becomes possible.
Again, one may need to use an extended domain to avoid discontinuity in $\dot \theta$ and use the extended domain $\theta \in [0,2\pi]$ for motions including $\theta=\pi$. 

The derivative of the effective potential, equation \eqref{ueff3}, can be obtained as
\begin{equation}
\frac{d U_{eff}(\theta)}{d \theta}= -\left[\frac{ I_x a^2 }{(1-\cos \theta)^2}+Mgl \right] \sin \theta.
\end{equation}
The terms in the parentheses can not be equal to zero since both terms are always positive.
This time, $\theta=\pi$ corresponds to the minimum of the effective potential with value $-Mgl$.
This situation can be seen in figure \ref{fig:ueff}(c).

Similar to the previous case by considering $E'=U_{eff}$ with $\dot \theta=0$ and changing variable, one can obtain a quadratic equation for $u$ as
\begin{equation} 
	u^2- \left(1+\frac{I_x a^2+2 E'}{2 Mgl} \right) u+ \frac{2 E' - I_x a^2}{2 Mgl}=0, \label{quadratic_aeqlmb}
\end{equation}
One of the roots of this equations gives turning angle, i.e. $\theta_{min}$, and the other one can be defined as $2 \pi-\theta_{min}$ for extended potential, see figure \ref{fig:ueff}(c). 

Angular velocities $\dot \phi$ and $\dot \psi$ can be found as \cite{Tanriverdi_abeql}
\begin{eqnarray}
\dot \phi &=& \frac{-a}{1-\cos \theta}, \\
\dot \psi &=& a\left( \frac{I_x}{I_z} +\frac{\cos \theta}{1-\cos \theta}\right).
\end{eqnarray}

\section{Small oscillations when conserved angular momenta are equal}
\label{two}

We will consider the small oscillations in two parts since there are two possible situations, i.e. $b=a$ and $b=-a$.

For examples, we will take $I_x=I_y=22.8 \times 10^{-5}\, kg\, m^2$, $I_z=5.72 \times 10^{-5}\, kg\, m^2$ and $Mgl=0.068 J$.
We will compare the results of the small oscillation approximation with the results of the numerical integration of angular accelerations, whose details can be found in a previous work \cite{Tanriverdi_abeql}.

\subsection{Small oscillations when $b=a$}

When $b=a$, equation \eqref{ddottheta} can be written as
\begin{equation}
\ddot \theta=-\frac{a^2}{\sin^3 \theta} (1-\cos \theta)^2+\frac{Mgl}{I_x} \sin \theta. \label{ddottheta2}
\end{equation}
For small oscillations, one can write $\theta=\theta_0+\eta$ and $\sin \theta \approx \sin \theta_0 + \eta \cos \theta_0$ and $\cos \theta \approx \cos \theta_0 - \eta \sin \theta_0$ where $\eta$ is small.
By using these in equation \eqref{ddottheta2}, expanding denominator and ignoring higher-order terms, one can obtain
\begin{equation}
	\ddot \eta + \eta \left[ \frac{a^2(2-\cos \theta_0)}{(1+ \cos \theta_0)^2}-\frac{Mgl}{I_x} \cos \theta_0 \right]\approx \frac{Mgl}{I_x} \sin \theta_0-\frac{a^2(1-\cos \theta_0)}{\sin \theta_0(1+ \cos \theta_0)}. \label{fho} 
\end{equation}
One can define $w_1^2=a^2(2-\cos \theta_0)/(1+ \cos \theta_0)^2-Mgl \cos \theta_0/I_x$ and $C_1=Mgl \sin \theta_0/I_x-a^2(1-\cos \theta_0)/[\sin \theta_0(1+ \cos \theta_0)]$.
Then, by considering $\ddot \eta + \eta w_1^2 \approx C_1$, $\dot \eta (t=0)=0$ and $\eta(t=0)=0$, one can obtain the solution as
\begin{equation}
	\eta(t)=\frac{C_1}{w_1^2}(1- \cos(w_1 t)).
\end{equation}
Then by considering $\theta= \theta_0+\eta$, one can obtain
\begin{eqnarray}
	\theta(t)&\approx&\theta_0+\frac{C_1}{w_1^2}(1- \cos(w_1 t)), \label{theta1}\\
	\dot \theta(t) &\approx& \frac{C_1}{w_1} \sin(w_1 t),  \label{dottheta1} 
\end{eqnarray}
and by using equations for $\dot \phi=(b- a \cos \theta)/\sin^2 \theta$ and $\dot \psi= I_x a/I_z-(b- a \cos \theta)\cos \theta /\sin^2 \theta $ and small oscillations, one can obtain
\begin{eqnarray}
	\dot \phi(t) &\approx& \dot \phi_0 +\frac{\dot \phi_0 (1-\cos \theta_0)}{\sin \theta_0}\frac{C_1}{w_1^2}(1- \cos(w_1 t)), \label{dotphi1} \\
	\dot \psi(t) &\approx& \dot \psi_0 +\frac{\dot \phi_0 (1-\cos \theta_0)}{\sin \theta_0}\frac{C_1}{w_1^2}(1- \cos(w_1 t)), \label{dotpsi1} \\
	\phi(t) &\approx& \dot \phi_0 t+\frac{\dot \phi_0 (1-\cos \theta_0)}{\sin \theta_0}\frac{C_1}{w_1^2}\left(t- \frac{\sin(w_1 t)}{w_1}\right), \label{phi1} \\
	\psi(t) &\approx& \dot \psi_0 t+\frac{\dot \phi_0 (1-\cos \theta_0)}{\sin \theta_0}\frac{C_1}{w_1^2}\left(t- \frac{\sin(w_1 t)}{w_1}\right). \label{psi1}
\end{eqnarray}
$\phi$ and $\psi$ are obtained by integrating corresponding angular velocities and considering $\phi(t=0)=0$ and $\psi(t=0)=0$.
These equations are adequate for small oscillations when the minimum of the effective potential is between $0$ and $\pi$.

\begin{figure}[h!]
\begin{center}
\subfigure[$\theta$]{
\includegraphics[width=4.4cm]{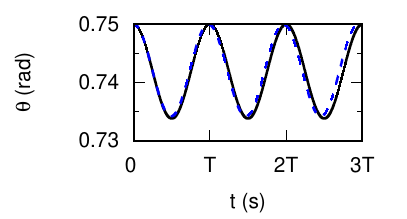}
}
\subfigure[$\dot \theta$]{
\includegraphics[width=4.4cm]{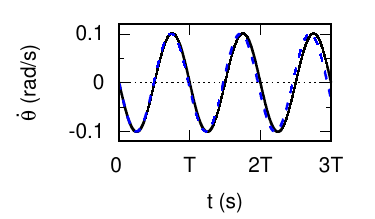}
}
\subfigure[$\dot \phi$]{
\includegraphics[width=4.4cm]{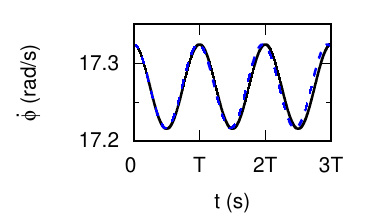}
}
\subfigure[$\dot \psi$]{
\includegraphics[width=4.4cm]{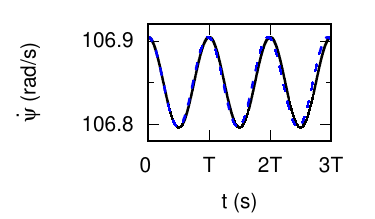}
}
\caption{
	Time evolution of $\theta$ (a), $\dot \theta$ (b), $\dot \phi$ (c) and $\dot \psi$ (d) for slow top ($b=a$).
        Continuous (black) curves show the results of numerical integration of angular accelerations, dashed (blue) curves show results for the small oscillations with $w_1$.
        Initial values are $\theta_0=0.75\, rad$, $\dot \theta_0=0$, $\dot \phi_0=17.3 \,rad\,s^{-1} $ and $\dot \psi_0=107 \,rad\,s^{-1}$, and $T=0.251\, s$.
        }
\label{fig:thetaphipsi_1}
\end{center}
\end{figure}

One can see an example of the small oscillation approximation in figure \ref{fig:thetaphipsi_1} for slow top whose minimum is between $0$ and $\pi$.
For this example, $b=a=30 \,rad\,s^{-1}$ which is smaller than $\sqrt{4 Mgl/I_x}$ and $E'=0.0656\,J$ and inital values can be found in the explanations of that figure.
From the figure, it can be seen that the small oscillation approximation gives close values to the results of the numerical integration of angular accelerations.
The angular frequency obtained from numerical integration of angular accelerations is $12.5 \,rad\,s^{-1}$, and it can be obtained by using formula given for $w_1$ after equation \eqref{fho} as $12.7 \,rad\,s^{-1}$.
The percentage difference for angular frequency is $1.8\%$, and it is $2.3\%$ for the amplitude of oscillations of $\theta$. 
The shapes for the locus for this case can be seen in figure \ref{fig:gr_tt}(a), and it can be seen that the small oscillation approximation, equations \eqref{theta1} and \eqref{phi1}, gives similar shapes for the locus to the numerical integration of angular accelerations.

As it is mentioned previously when the minimum of the effective potential is at $\theta=0$, one may need to use the extended domain, i.e. $\theta \in [-\pi,\pi]$.
$\theta$ can take negative values during the usage of the extended domain, and these negative values show the same inclinations with positive values,
which means the real nutation period of the motion is half of the one obtained by considering the extended domain.
Then, angular frequency $w_1$ should be double for motions which includes $\theta=0$.
In these cases, equations \eqref{theta1} and \eqref{dottheta1} can still be used to find $\theta$ and $\dot \theta$, however,
for $\dot \phi$, $\dot \psi$, $\phi$ and $\psi$, one need to use equations (\ref{dotphi1}-\ref{psi1}) by writing $2 w_1$ instead of $w_1$.

\begin{figure}[h!]
\begin{center}
\subfigure[$\theta$]{
\includegraphics[width=4.4cm]{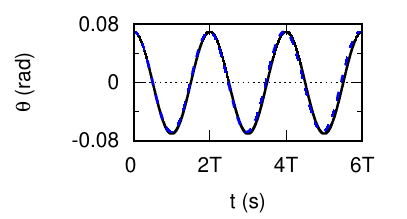}
}
\subfigure[$\dot \theta$]{
\includegraphics[width=4.4cm]{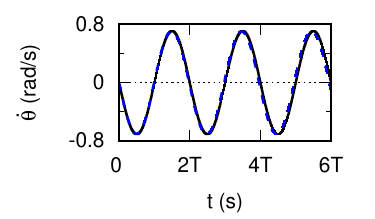}
}
\subfigure[$\dot \phi$]{
\includegraphics[width=4.4cm]{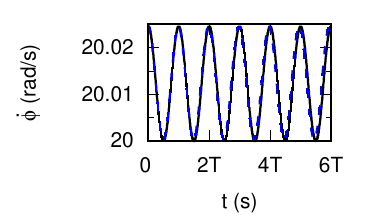}
}
\subfigure[$\dot \psi$]{
\includegraphics[width=4.4cm]{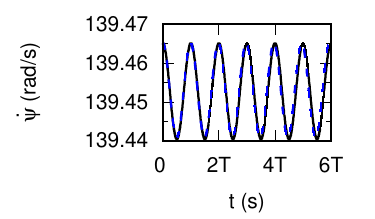}
}
\caption{
	Time evolution of $\theta$ (a), $\dot \theta$ (b), $\dot \phi$ (c) and $\dot \psi$ (d) for fast top ($b=a$).
        Continuous (black) curves show the results of numerical integration of angular accelerations, dashed (blue) curves show results for the small oscillations with $w_1$.
	For the small oscillation approximation, $w_1$ is replaced with $2 w_1$ in equations \eqref{dotphi1} and \eqref{dotpsi1}.
        Initial values are $\theta_0=0.07\, rad$, $\dot \theta_0=0$, $\dot \phi_0=20.0 \,rad\,s^{-1} $ and $\dot \psi_0=139 \,rad\,s^{-1}$, and $T=0.310\, s$ without considering extended domain.
        }
\label{fig:thetaphipsi_2}
\end{center}
\end{figure}

An example of this situation can be seen in figure \ref{fig:thetaphipsi_2}.
For this example, $b=a=40 \,rad\,s^{-1}$ which is greater than $\sqrt{4 Mgl/I_x}$, and $E'=0.0681\,J$.
From the figure, it can be seen that approximate solutions can be obtained by considering small oscillations.
In this case, the percentage difference for angular frequency can be obtained as $1.0\%$ by using angular frequencies $w=10.1 \,rad\,s^{-1}$ calculated by considering $2T$ and $w_1=10.2 \,rad\,s^{-1}$ calculated by given formula for $w_1$,
and the percentage difference for the amplitude of $\theta$ is $1.7\%$
For this case, the shapes for the locus can be seen in figure \ref{fig:gr_tt}, while calculating $\phi$ from the small oscillation approximation $w_1$ is replaced with $2 w_1$.
And, it can be seen that the shapes for the locus can also be obtained from this approximation.

\begin{figure}[h!]
\begin{center}
\subfigure[]{
\includegraphics[width=4.0cm]{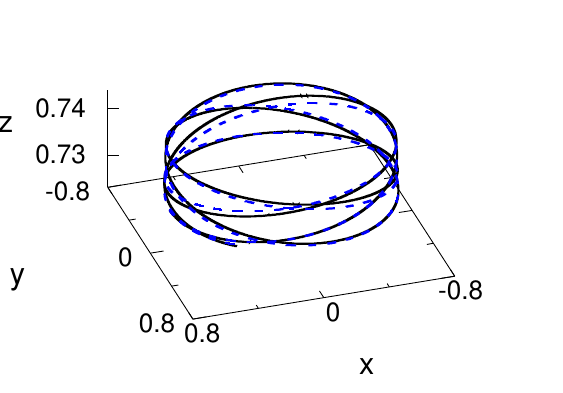}
}
\subfigure[]{
\includegraphics[width=4.0cm]{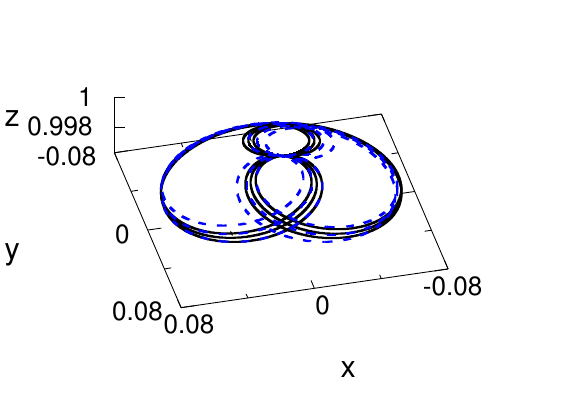}
}
\subfigure[]{
\includegraphics[width=4.0cm]{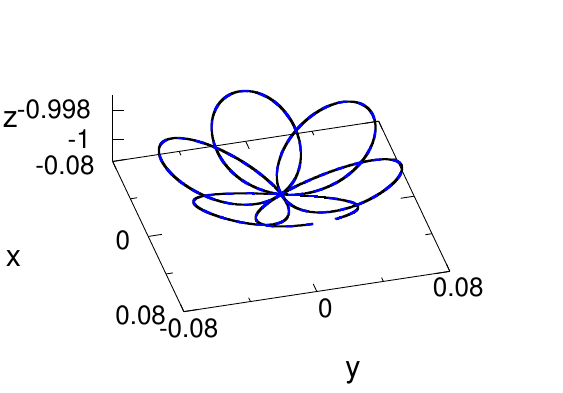}
}
\caption{
        Shapes for the locus. Continuous (black) curves show results of numerical integration of angular accelerations, 
	and dashed (blue) curves show the results of the small oscillation approximation.
	(a) Slow top when $b=a$, initial values are given in figure \ref{fig:thetaphipsi_1}. The animated version can be found at \href{https://youtu.be/AzKI1E-VsAE}{https://youtu.be/AzKI1E-VsAE}.
	(b) Fast top when $b=a$, initial values are given in figure \ref{fig:thetaphipsi_2}. The animated version can be found at \href{https://youtu.be/jepQpyHt-qI}{https://youtu.be/jepQpyHt-qI}.
	(c) Motion $b=-a$, initial values are given in figure \ref{fig:thetaphipsi_3}. The animated version can be found at \href{https://youtu.be/Io8d\_gndZD8}{https://youtu.be/Io8d\_gndZD8}.
        }
\label{fig:gr_tt}
\end{center}
\end{figure}

\subsection{Small oscillations when $b=-a$}

When $b=-a$, equation \eqref{ddottheta} can be written as
\begin{equation}
\ddot \theta=\frac{a^2}{\sin^3 \theta} (1+\cos \theta)^2+\frac{Mgl}{I_x} \sin \theta . \label{ddottheta3}
\end{equation}
Similar to the previous case from equation \eqref{ddottheta3}, one can obtain
\begin{equation}
	\ddot \eta + \eta \left[ \frac{a^2(2+\cos \theta_0)}{(1- \cos \theta_0)^2}-\frac{Mgl}{I_x} \cos \theta_0 \right]\approx \frac{Mgl}{I_x} \sin \theta_0+\frac{a^2(1+\cos \theta_0)}{\sin \theta_0(1- \cos \theta_0)}. \label{sho}
\end{equation}
One can define $w_2^2=a^2(2+\cos \theta_0)/(1- \cos \theta_0)^2-Mgl \cos \theta_0/I_x$ and $C_2=Mgl \sin \theta_0/I_x+a^2(1+\cos \theta_0)/[\sin \theta_0(1- \cos \theta_0)]$.
Then, by considering $\ddot \eta + \eta w_2^2 \approx C_2$, $\dot \eta (t=0)=0$ and $\eta(t=0)=0$, one can obtain 
\begin{equation}
        \eta(t)=\frac{C_2}{w_2^2}(1- \cos(w_2 t)).
\end{equation}
Then, similar to previous case, by considering $\theta= \theta_0+\eta$, one can obtain
\begin{eqnarray}
	\theta(t)&\approx&\theta_0+\frac{C_2}{w_2^2}(1- \cos(w_2 t)), \label{theta3} \\
	\dot \theta(t) &\approx& \frac{C_2}{w_2} \sin(w_2 t),  \label{dottheta3}
\end{eqnarray}	
and similarly by considering equations $\dot \phi$ and $\dot \psi$ and small oscillations, one can obtain
\begin{eqnarray}
	\dot \phi(t) &\approx& \dot \phi_0 -\frac{\dot \phi_0 (1+\cos \theta_0)}{\sin \theta_0}\frac{C_2}{w_2^2}(1- \cos(w_2 t)), \label{dotphi34} \\
        \dot \psi(t) &\approx& \dot \psi_0 +\frac{\dot \phi_0 (1+\cos \theta_0)}{\sin \theta_0}\frac{C_2}{w_2^2}(1- \cos(w_2 t)), \label{dotpsi34} \\
        \phi(t) &\approx& \dot \phi_0 t-\frac{\dot \phi_0 (1+\cos \theta_0)}{\sin \theta_0}\frac{C_2}{w_2^2}\left(t- \frac{\sin(w_2 t)}{w_2}\right), \label{phi34} \\
        \psi(t) &\approx& \dot \psi_0 t+\frac{\dot \phi_0 (1+\cos \theta_0)}{\sin \theta_0}\frac{C_2}{w_2^2}\left(t- \frac{\sin(w_2 t)}{w_2}\right), \label{psi34}
\end{eqnarray}
where $\phi(t=0)=0$ and $\psi(t=0)=0$.

When $b=-a$, the minimum of the effective potential is always at $\theta=\pi$.
Then, the motion includes the point $\theta=\pi$, and one needs to use the extended domain, i.e. $\theta\in[0,2\pi]$.
Similar to the previous usage of the extended domain, one needs to multiply angular frequency with $2$ and write $2 w_2$ instead of $w_2$ in equations (\ref{dotphi34}-\ref{psi34}).

\begin{figure}[h!]
\begin{center}
\subfigure[$\theta$]{
\includegraphics[width=4.4cm]{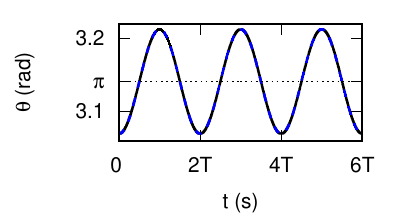}
}
\subfigure[$\dot \theta$]{
\includegraphics[width=4.4cm]{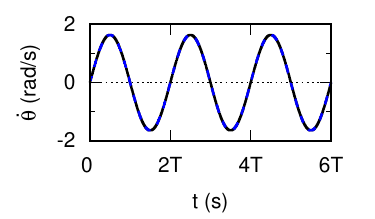}
}
\subfigure[$\dot \phi$]{
\includegraphics[width=4.4cm]{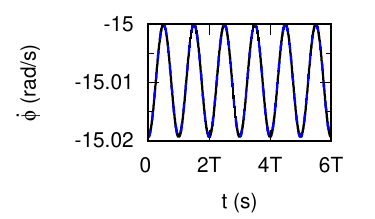}
}
\subfigure[$\dot \psi$]{
\includegraphics[width=4.4cm]{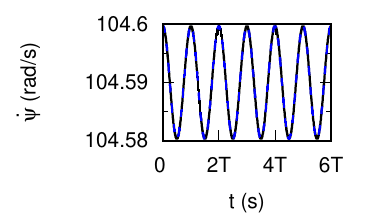}
}
\caption{
Time evolution of $\theta$ (a), $\dot \theta$ (b), $\dot \phi$ (c) and $\dot \psi$ (d) when $b=-a$.
        Continuous (black) curves show the results of numerical integration of angular accelerations, dashed (blue) curves show results for the small oscillations with $w_2$.
        Initial values are $\theta_0=3.07\, rad$, $\dot \theta_0=0$, $\dot \phi_0=-15.0 \,rad\,s^{-1} $ and $\dot \psi_0=105 \,rad\,s^{-1}$, and $T=0.137\, s$.
        }
\label{fig:thetaphipsi_3}
\end{center}
\end{figure}

An example of this case can be seen in figure \ref{fig:thetaphipsi_3}.
For this case, $-b=a=30.0 \,rad\,s^{-1}$ and $E'=-0.0677\,J$, and initial values can be found in the explanations of the figure.
As it can be seen from the figure, the small oscillation approximation gives very close results to the numerical integration of angular accelerations.
Angular frequencies can be obtained as $w=22.87 \,rad\,s^{-1}$ from numerical integration and $w_2=22.88 \,rad\,s^{-1}$ from the relation given after equation \eqref{sho}, 
and their percentage difference is $0.03 \%$.
It is $0.05\%$ for the amplitude of oscillation of $\theta$.
The shapes for locus can be seen in figure \ref{fig:gr_tt}(c).

\section{Conclusion}
\label{three}

We have studied small oscillations of a heavy symmetric top when conserved angular momenta are equal to each other, i.e. $|b|=|a|$.
The small oscillation approximation is done by using the angular acceleration for inclination angle $\ddot \theta$ when $b=a$ and $b=-a$ ($Mgl>0$).
Overall results are consistent with the numerical solution.

In the appendix, we studied the situation when $Mgl$ is negative.
The results have shown that the small oscillation approximation can also be used in this case.

To obtain results closer to the exact one, it is needed to choose $E'$ values very close to the minimum of the effective potential.
As the difference between $E'$ and $U_{eff_{min}}$ becomes greater, the similarity between the results of the small oscillation approximation and the numerical solution disappears as expected.

\section{Appendix}

In this part, we will consider the case with negative $Mgl$ which changes the effective potential.
This situation is possible for some symmetric tops with counterweight.

\begin{figure}[h!]
\begin{center}
        \subfigure[$b=a$ ]{
\includegraphics[width=4.0cm]{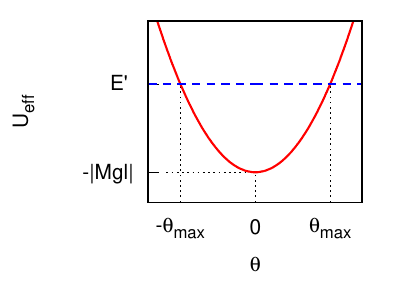}
}
\subfigure[$b=-a$ (slow top)]{
\includegraphics[width=4.0cm]{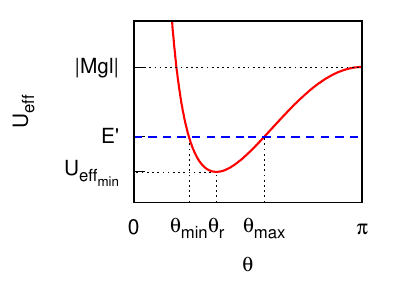}
}
\subfigure[$b=-a$ (fast top)]{
\includegraphics[width=4.0cm]{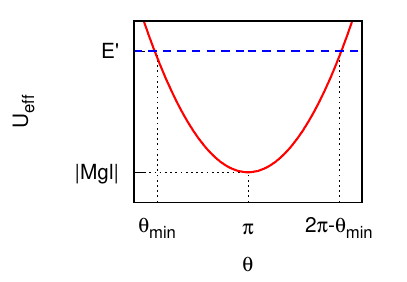}
}
\caption{
        Effective potentials when $Mgl<0$.
        (a) When $b=a$
        (b) For slow top when $b=-a$
        (c) For fast top when $b=-a$
        }
\label{fig:ueff2}
\end{center}
\end{figure}

When $b=a$, the minimum of the effective potential is always at $\theta=0$ for fast and slow tops, and its value is equal to $-|Mgl|$.
The turning angle can be found by using equation \eqref{quadratic_abeql} and writing $-|Mgl|$ instead of $Mgl$, 
and the other turning angle can be considered as $-\theta_{max}$ for the extended domain, see figure \ref{fig:ueff2}(a).

When $b=-a$, there are two possible situations.
In the first case, when $|a|<\sqrt{4|Mgl|/I_x}$, i.e. slow top, $\theta=\pi$ corresponds to the local maximum with value $|Mgl|$.
The minimum of the effective potential is at $\theta_r =\arccos(1-\sqrt{I_x a^2/|Mgl|})$ with value $U_{eff_{min}}=2\sqrt{|Mgl| I_x a^2}-|Mgl|-I_x a^2/2$.
And, there are two turning angles between $0$ and $\pi$ which can be easily found by writing $Mgl=-|Mgl|$ in equation \eqref{quadratic_aeqlmb}.
Such a case can be seen in figure \ref{fig:ueff2}(b).

In the second case, when $|a|>\sqrt{4|Mgl|/I_x}$, i.e. fast top, $\theta=\pi$ corresponds to the minimum with the value $|Mgl|$.
And, there is only one turning angle $\theta_{min}$ which can be found by using equation \eqref{quadratic_aeqlmb} and writing $Mgl=-|Mgl|$ at that equation, again.
The other turning angle can be considered as $2\pi-\theta_{min}$, see figure \ref{fig:ueff2}(c).

\begin{figure}[h!]
\begin{center}
\subfigure[$b=a$]{
\includegraphics[width=4.0cm]{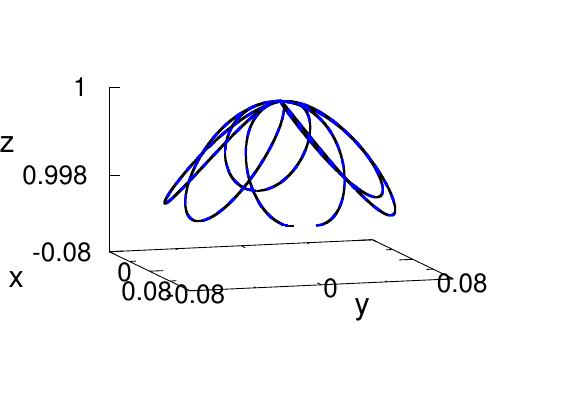}
}
\subfigure[$b=-a$ (slow top)]{
\includegraphics[width=4.0cm]{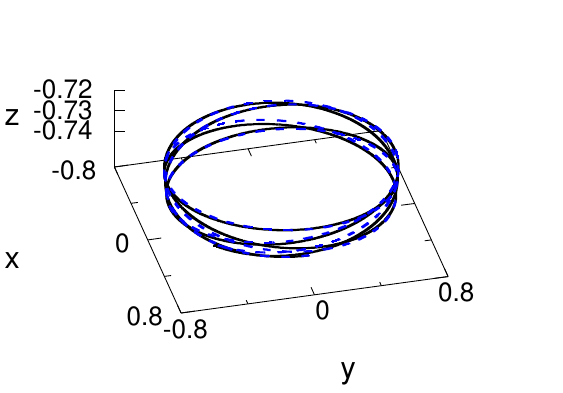}
}
\subfigure[$b=-a$ (fast top)]{
\includegraphics[width=4.0cm]{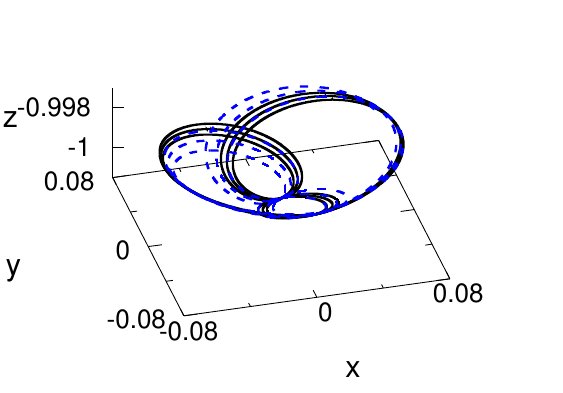}
}
\caption{
        Shapes for the locus when $Mgl<0$. Continuous (black) curves show results of numerical integration of angular accelerations,
        and dashed (blue) curves show the results of the small oscillation approximation.
        (a) Motion when $b=a$, initial values are $\theta_0=0.07\, rad$, $\dot \theta_0=0$, $\dot \phi_0=15.0 \,rad\,s^{-1} $ and $\dot \psi_0=105 \,rad\,s^{-1}$, and $T=0.275 \, s$. 
        (b) Slow top when $b=-a$, initial values are $\theta_0=2.41\, rad$, $\dot \theta_0=0$, $\dot \phi_0=-17.2 \,rad\,s^{-1} $ and $\dot \psi_0=107 \,rad\,s^{-1}$, and $T=0.502 \, s$.
        (c) Fast top when $b=-a$, initial values are $\theta_0=3.07\, rad$, $\dot \theta_0=0$, $\dot \phi_0=-20.0 \,rad\,s^{-1} $ and $\dot \psi_0=139 \,rad\,s^{-1}$, and $T=0.621 \, s$.
        }
\label{fig:gr_tt2}
\end{center}
\end{figure}

Now, we will consider three examples, and for all of them $Mgl=-0.068\,J$.
Initial values for all three cases can be found in the explanations of figure \ref{fig:gr_tt2}.
We will show the plots of the shapes for the locus for examples and will not show the plots of changes in $\theta$ and angular velocities since they are not very different than previous cases.

The first one is an example of the small oscillation approximation when $b=a$.
For this case, $a=30 \,rad\,s^{-1}$ and $E'=-0.0677\,J$ which is slightly bigger than the minimum of effective potential which is equal to $-|Mgl|$.
One can see the shapes for the locus obtained from the results of the small oscillation approximation and numerical integration of angular accelerations in figure \ref{fig:gr_tt2}(a).
For the small oscillation approximation, $\theta$ is obtained from equation \eqref{theta1}, and $\phi$ is obtained from equation \eqref{phi1} by writing $2 w_1$ instead of $w_1$.

For the second example, a case with $|a|<\sqrt{4 Mgl /I_x}$ and $b=-a$ is chosen, and oscillations are around the minimum which is between $0$ and $\pi$.
For this example, $-b=a=30.0 \,rad\,s^{-1}$ and $E'=0.065653 \,J$, and the minimum occurs at $\theta=2.40\,rad$ with value $U_{eff_{min}}=0.065651 \,J$.
It can be seen from the shapes for the locus that the numerical integration and the small oscillation approximation give similar results, see figure \ref{fig:gr_tt2}(b).

The third example is a fast top when $-b=a=40.0 \,rad\,s^{-1}$ and $E'=0.0681 \,J$.
In this case, the minimum of the effective potential occurs at $\theta=\pi$ and $U_{eff_{min}}=|Mgl|$, and we will use the extended domain.
Then, equations \eqref{theta3} and \eqref{dottheta3} can be used, however, equations (\ref{dotphi34}-\ref{psi34}) should be used by writing $2 w_2$ instead of $w_2$.
The results for the small oscillation approximation and numerical integration of angular accelerations for the shapes for the locus can be seen in figure \ref{fig:gr_tt2}(c).
It can be seen that the small oscillation approximation gives similar results to the numerical integration.

\end{document}